\documentclass[12pt]{iopart}
\usepackage{graphicx}
\usepackage{color}
\usepackage{bm}
\usepackage{ams}
\usepackage{amssymb}

\def\muSR{$\mu$SR}
\def\TN{$T_{\rm N}$}
\def\Tc{$T_{\rm c}$}
\def\chiQw{\ensuremath{\chi''(\bm{Q},E)}}

\def\fetesex{Fe$_{1+y}$Se$_x$Te$_{1-x}$}

\def\feteseA{Fe$_{1.01}$Se$_{0.50}$Te$_{0.50}$}
\def\feteseB{Fe$_{1.10}$Se$_{0.25}$Te$_{0.75}$}

\begin{document}
\title{Magnetic excitations of \fetesex \ in magnetic and superconductive phases}
\author{P. Babkevich$^{1,2}$, M. Bendele$^{3,4}$,
A. T. Boothroyd$^1$, K. Conder$^5$,
S. N. Gvasaliya$^2$, R. Khasanov$^3$,
E. Pomjakushina$^5$ and B.~Roessli$^2$}
\address{$ˆ1$ Department of Physics, Clarendon Laboratory,
Oxford University, Oxford OX1 3PU, United Kingdom}
\address{$ˆ2$ Laboratory for Neutron Scattering, Paul Scherrer Institut, CH-5232 Villigen PSI, Switzerland}
\address{$^3$ Laboratory for Muon Spin Spectroscopy, Paul Scherrer Institut, CH-5232 Villigen PSI, Switzerland}
\address{$^4$ Physik-Institut der Universit\"{a}t Z\"{u}rich, Winterthurerstrasse 190, CH-8057 Z\"{u}rich, Switzerland}
\address{$^5$ Laboratory for Developments and Methods, Paul Scherrer Institut, CH-5232 Villigen PSI, Switzerland}
\eads{peter.babkevich@physics.ox.ac.uk}

\begin{abstract}
We have used inelastic neutron scattering and muon-spin rotation to compare the
low energy magnetic excitations in single crystals of
superconducting \feteseA\ and non-superconducting \feteseB. We
confirm the existence of a spin resonance in the superconducting
phase of \feteseA, at an energy of 7\,meV and a wavevector of
$(1/2,1/2,0)$. The non-superconducting sample exhibits two
incommensurate magnetic excitations at
$(1/2,1/2,0)\pm(0.18,-0.18,0)$ which rise steeply in energy, but no
resonance is observed at low energies. A strongly dispersive
low-energy magnetic excitation is also observed in \feteseB\ close
to the commensurate antiferromagnetic ordering wavevector
$(1/2-\delta,0,1/2)$ where $\delta \approx 0.03$. The magnetic
correlations in both samples are found to be quasi-two dimensional
in character and persist well above the magnetic (\feteseB) and
superconducting (\feteseA) transition temperatures.
\end{abstract}

\maketitle

% Introduction ===============================================================
\section{Introduction}
Considerable effort has been devoted over the past two years to
investigate the basic properties of the Fe-based family of
superconductors \cite{ishida-jpsj-2009,lynn-physicac-2009,norman-phys-2008}.
%\cite{fang-prb-2008,hsu-pnas-2008,rotter-prl-2008,bao-prl-2009,qui-prl-2009,khasanov-prb-2009}.
A central question is whether magnetism plays an important role in
the formation of the superconducting state. A useful strategy to
tackle this problem is to combine neutron scattering and muon spin
rotation (\muSR) measurements on one and the same sample. Neutron
scattering provides information on magnetic correlations and on the
nature of the magnetic excitations, while \muSR\ can determine
whether static magnetic order and/or bulk superconductivity exists.

The \fetesex\ system is a convenient one to study with this
methodology as large high-quality crystals can be grown \cite{sales-prb-2009,chen-prb-2009} and the
tetragonal crystallographic structure is relatively simple to
analyse and model. Single crystals are easier to grow if there is a small excess of Fe (i.e. $y>0$), especially for small $x$ \cite{hsu-pnas-2008,gronwold-1954,bao-prl-2009,yeh-cgd-2009}.

The pure FeSe compound is a superconductor with a
transition temperature $T_{\rm c} \sim 8$\,K \cite{hsu-pnas-2008}. The $T_{\rm c}$ can be increased by
partial substitution of Te for Se such that $T_{\rm c} \sim 14$\,K for $0.4
\lesssim x \lesssim 0.8$ and $y \approx 0$
\cite{fang-prb-2008,khasanov-prb-2009}. The application of pressure
has also been found to raise \Tc, with values as high as 37\,K
observed for FeSe \cite{margadonna-prb-2009,mizuguchi-apl-2008,
medvedev-nature-2009,garbarino-epl-2009}. Compounds with $x
\lesssim 0.4$ do not exhibit bulk superconductivity but order
magnetically below a temperature which has a maximum of 67\,K at
$x=0$ and which decreases with $x$ and vanishes at $x \sim 0.4$. We
recently found evidence for coexistence of incommensurate magnetic
order and partial superconductivity for $x \sim 0.25$ \cite{khasanov-prb-2009}.

In this study, we present neutron scattering and \muSR\ measurements
on two single crystal samples: (i) \feteseA, a bulk superconductor,
and (ii) \feteseB, a non-superconducting sample which exhibits
magnetic order.  In superconducting \feteseA\ we observe a resonant
magnetic excitation consistent with that reported previously in
compounds with similar compositions
\cite{qui-prl-2009,iikubo-jpsj-2009,argyriou-preprint-2009,mook-preprint-2009v2}. In \feteseB\ we observe
strongly dispersive, low energy magnetic excitations associated with
the magnetic ordering wavevector $(1/2-\delta,0,1/2)$, $\delta
\approx 0.03$, and also with the wavevectors
$(1/2,1/2,0)\pm(0.18,-0.18,0)$. We find no evidence for a resonance
peak in the excitation spectrum of \feteseB.

% Experimental Description ===================================================
\section{Experimental Details}
Single crystals of \fetesex\ were grown by a modified Bridgman
method as reported by \cite{sales-prb-2009}. Neutron scattering measurements were carried out on triple-axis
spectrometer TASP \cite{semadeni-physicab-2001} at the SINQ \cite{fischer-physicab-1997} spallation source (Paul Scherrer
Institut, Switzerland). Bragg reflections from
pyrolytic graphite PG(002) monochromator and analyser were used at a
fixed final wavevector of 2.66\,\AA$^{-1}$. A PG filter was placed
after the sample to reduce contamination from higher order harmonics
in the beam and the instrument was set up in the open -- open --
open -- open collimation with the analyser focusing in the
horizontal plane. The crystals were single rods with masses of
approximately 4\,g. The \feteseB \ sample was orientated in two
settings to give access to $(h,0,l)$ and $(h,k,0)$ planes in
reciprocal space. Measurements of \feteseA\ were made in the
$(h,k,0)$ plane only. In this report we index
the reciprocal lattice with respect to the primitive tetragonal unit
cell described by the $P4/nmm$ space group with unit cell parameters
$a \approx 3.8$\,\AA\ and $c \approx 6.1$\,\AA\ along lines joining
the nearest neighbour Fe atoms.

Zero-field-cooled magnetisation measurements were
performed on a Quantum Design MPMS magnetometer with a measuring
field $\mu_0H=0.3$\,mT using the direct current method. To reduce the effects of demagnetisation,
thin plate-like pieces of \fetesex, cleaved from
the main single crystals, were oriented with the flat surface ($ab$
plane) parallel to the applied field.

Zero-field (ZF) and transverse-field (TF) muon-spin rotation
(\muSR) experiments were performed on the $\pi$M3 beam line at
S$\mu$S (Paul Scherrer Institut, Switzerland). In TF experiments a
magnetic field of 11.8\,mT was applied parallel to the
crystallographic $ab$ plane of the crystal and perpendicular to the
muon-spin polarisation.

% Results ====================================================================
\section{Results}

\subsection{Magnetisation and \muSR\ Measurements}\label{sec:muSR}

\begin{figure}[htb]
\includegraphics[width=0.9\textwidth]{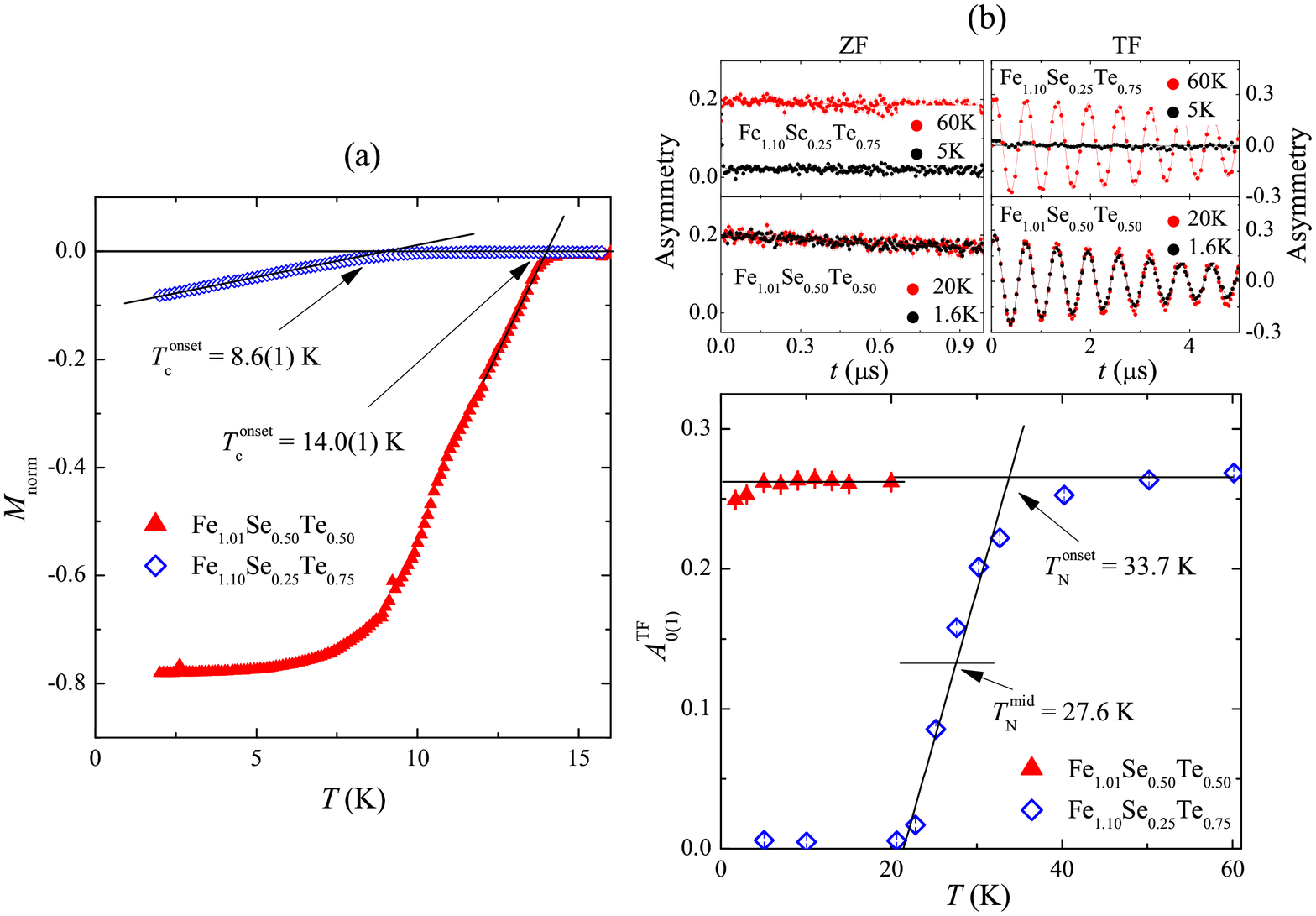}
 %\vspace{-0.5cm}
\caption{(a) Temperature dependence of the
zero-field-cooled magnetisation of \feteseA\ and \feteseB normalised to the ideal $1/4\pi$ value. The onset $T_{\rm c}^{\rm onset}$ of the superconducting transition is determined from
the intersection of straight lines fit to the data above and below
the transition. (b) Representative ZF and TF $\mu$SR time
spectra (upper plots) and temperature-dependent initial TF asymmetry
of the slow relaxing component ($A_0^{\rm TF}$ and $A_1^{\rm TF}$,
lower plot) for single crystals of \feteseA\ and \feteseB. The onset
($T_{\rm N}^{\rm onset}$) and the mid-point ($T_{\rm N}^{\rm mid}$)
of the magnetic transition are determined from the intersection of
straight lines fit to the data above and below the transition and as
the point where the asymmetry decreases by a factor of 2 from its
maximum value, respectively. }
 \label{fig:muSR}
\end{figure}

Zero-field-cooled magnetisation data normalised to the ideal
$1/4\pi$ value ($M_{\rm norm}$) are shown in
figure~\ref{fig:muSR}(a).
The \feteseA\ sample is seen to be a bulk
superconductor with the onset of the transition $T_{\rm c}^{\rm
onset}\simeq 14.0$\,K and $M_{\rm norm}\simeq-0.8$ at $T\simeq2$\,K.
The \feteseB\ sample also exhibits
superconductivity ($T_{\rm c}^{\rm onset}\simeq 8.6$\,K) but has a
small superconducting fraction of order 10\% at low temperature.

For the $x=0.5$ sample,
the ZF time-spectra measured at $T=1.7$\,K and 20\,K are almost
identical thus suggesting that the magnetic state of this sample is
the same above and below the superconducting transition temperature.
The solid lines correspond to a fit by the function $A^{\rm
ZF}(t)=A_0^{\rm ZF} e^{-\Lambda^{\rm ZF} t}$, where $A_0^{\rm ZF}$
is the initial asymmetry and $\Lambda^{\rm ZF}$ is the exponential
relaxation rate. Such behavior is consistent with dilute Fe moments
as observed recently for FeSe$_{1-x}$ \cite{khasanov-prb-2008}. The
TF data for $x=0.5$ fit well to the function $A^{\rm TF}(t)=A_0^{\rm
TF}e^{-(\Lambda^{\rm TF} t+\sigma^2t^2)}\cos(\gamma_\mu Bt+\phi)$.
Here, $\gamma_\mu/2\pi= 135.5$\,MHz/T is the muon gyromagnetic ratio,
$\phi$ is the initial phase of the muon-spin ensemble, and $\sigma$
is the Gaussian relaxation rate. Figure~\ref{fig:muSR}(b) shows that the TF asymmetry $A_0^{\rm TF}$ is
almost temperature independent.  The slightly stronger relaxation of
the muon-spin polarisation at 1.7\,K relative to 20\,K is due to
formation of the vortex lattice at $T<T_{\rm c}$.

Static (within the $\mu$SR time window) magnetism develops in
\feteseB\ as signalled by a fast drop of both
$A^{\rm ZF}$ and $A^{\rm TF}$ within the first 100\,ns (see upper panel of figure~\ref{fig:muSR}(b)). The solid lines correspond to fits with
$A^{\rm ZF}(t)=A_1^{\rm ZF} e^{-\Lambda_1^{\rm ZF} t}+A_2^{\rm ZF}
e^{-\Lambda_2^{\rm ZF} t}$ and $A^{\rm
TF}(t)=e^{-\sigma^2t^2/2}[A_1^{\rm TF}e^{-\Lambda_1^{\rm TF}
t}\cos(\gamma_\mu B_1t+\phi)+A_2^{\rm TF}e^{-\Lambda_2^{\rm TF}
t}\cos(\gamma_\mu B_2t+\phi)]$. Here, $A_{1(2)}^{\rm ZF(TF)}$ and
$\Lambda_{1(2)}^{\rm ZF(TF)}$ are the initial ZF (TF) asymmetry and
the exponential depolarisation rate of the slow (fast) relaxing
component, respectively.
The temperature evolution of $A_{1}^{\rm TF}$, shown in figure~\ref{fig:muSR}(b), reveals that below 20\,K magnetism occupies more than 95\% of
the whole sample volume. The corresponding values of the onset and
the mid-point of the magnetic transitions, determined as shown in
the figure, are $T_{\rm N}^{\rm onset}\simeq 33.7$\,K and $T_{\rm
N}^{\rm mid} \simeq 27.6$\,K. We note that although the magnetic order is shown to extend throughout virtually the entire volume of the sample, \muSR\ cannot determine whether the magnetic order is long range. The neutron diffraction data presented in the next section show that the magnetic order is in fact relatively short range.

\subsection{Neutron Scattering Results}\label{sec:ins}

\begin{figure}
\centering
\includegraphics[width=0.8\textwidth]{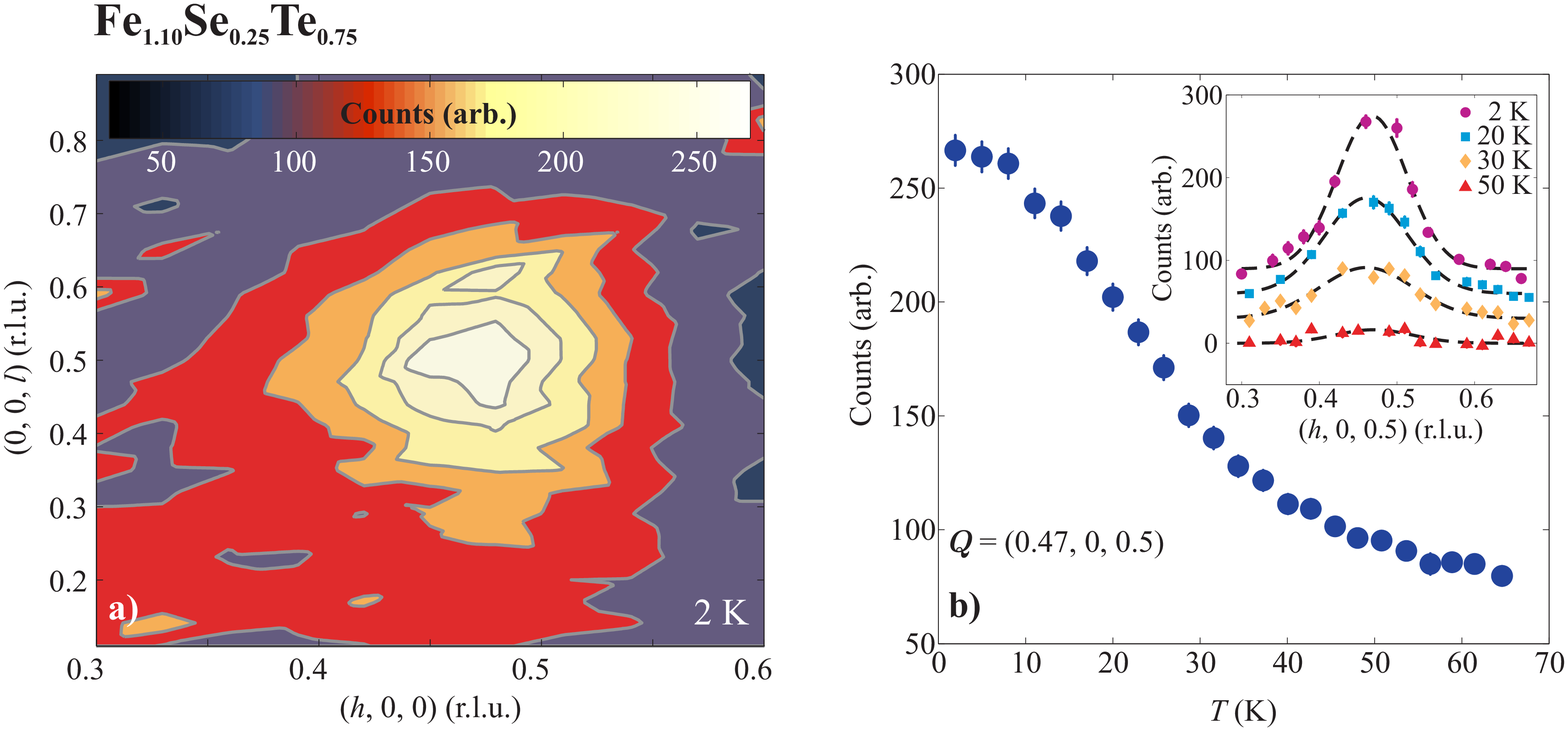}
\caption{Elastic neutron measurements of \feteseB\ at the magnetic
order propagation vector $\bm{q} = (0.47,0,0.5)$. (a) Map showing
the incommensurate peak at $\bm{q}$ in the $(h,0,l)$ plane at 2\,K.
(b) Temperature dependence of the intensity at $\bm{q}$. Inset
shows scans along $(h,0,0.5)$ measured at 2, 20, 30 and 50\,K. A
sloping background function has been subtracted from the data and
the dashed lines show a Gaussian fit through the peaks. For clarity,
the scans have been displaced vertically.}\label{fig:fetese19:1}
\end{figure}

% elastic measurements
Elastic neutron scattering measurements on \feteseB\ in the
$(h,0,l)$ scattering plane at 2\,K, as shown in
figure~\ref{fig:fetese19:1}(a), reveal a diffuse magnetic peak
centred on $(1/2-\delta,0,1/2)$ with $\delta \approx 0.03$. The
incommensurate peak is much broader than the resolution of the
instrument. From $\bm{Q}$ scans through the peak we obtain
correlation lengths along $a$ and $c$ axes of 11.4(6)\,\AA\ and
7.5(4)\,\AA\ respectively at 2\,K. Figure~\ref{fig:fetese19:1}(b)
shows that the magnetic peak develops below $T_{\rm N} \sim50$\,K. The
correlation lengths did not change measurably upon warming through
the \TN\ (figure~\ref{fig:fetese19:1}(b): inset). The magnetic
propagation vector $\bm{q}=(1/2-\delta,0,\pm1/2)$ is similar to that
found previously for the similar composition
Fe$_{1.03}$Se$_{0.25}$Te$_{0.75}$ compound. For the latter compound
we confirmed that the peak described by $\bm{q}$ is magnetic in
character using neutron polarisation analysis
\cite{khasanov-prb-2009}. Our results are consistent with measurements on Fe$_{1.07}$Se$_{0.25}$Te$_{0.75}$ for which the incommensurability is found to be $\delta \approx 0.04$ \cite{wen-prb-2009}.

\begin{figure}
\centering
\includegraphics[width=0.8\textwidth]{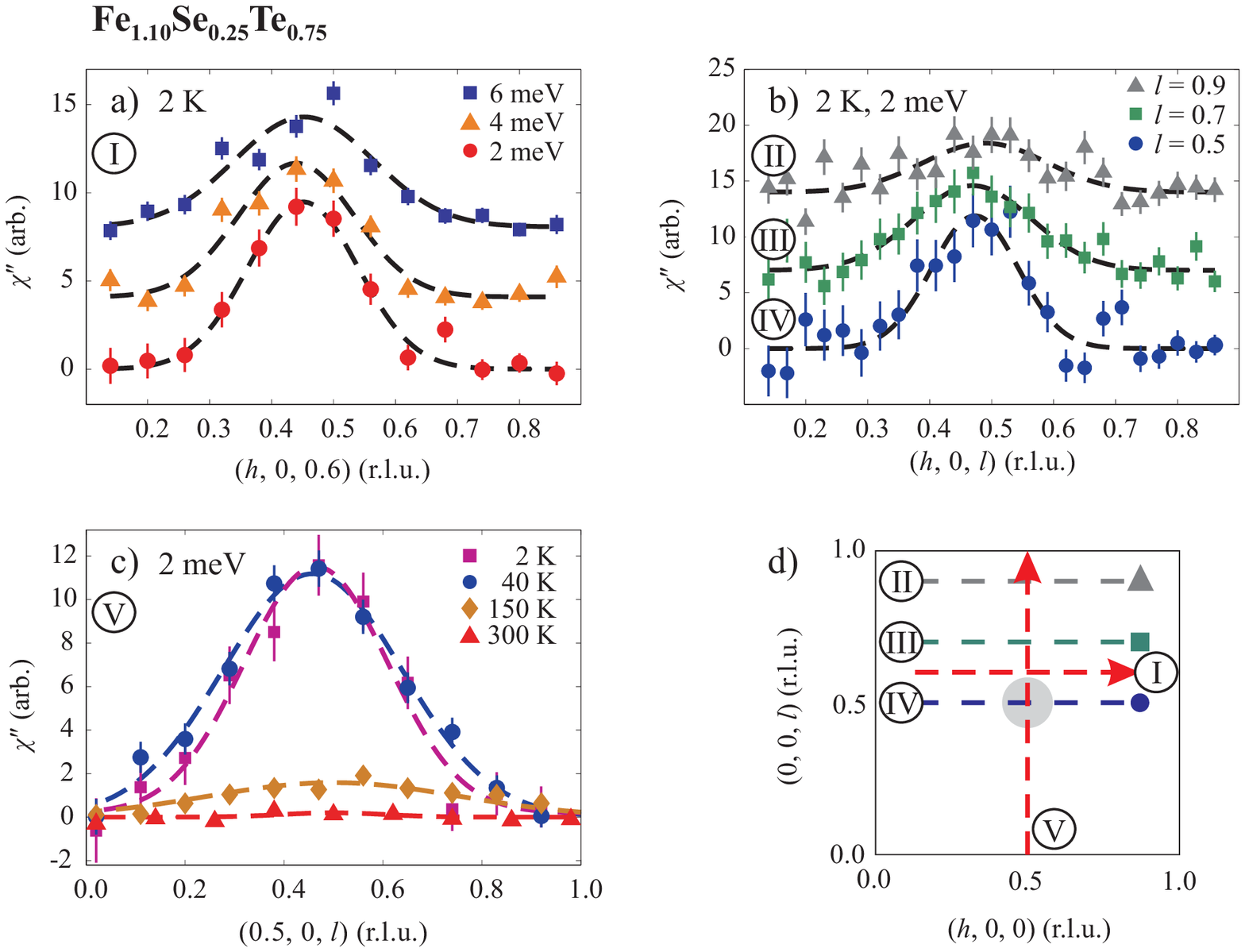}
\caption{Inelastic neutron scattering from \feteseB\ in the vicinity of the magnetic ordering wavevector $\bm{q}=(0.47,0,0.5)$. (a) Constant energy scans collected at
2, 4 and 6\,meV and 2\,K along $(h,0,0.6)$. The data has been shifted
in \chiQw\ by arbitrary amounts for clarity. (b) Constant energy
scans collected at 2\,meV and temperature of 2\,K showing \chiQw\
along $(h,0,0.5)$, $(h,0,0.7)$ and $(h,0,0.9)$. The plots have been
displaced and the dashed lines show Gaussian peaks through the
spectra. (c) Constant energy scans at 2\,meV at temperatures of 2, 40, 150 and 300\,K showing \chiQw\ along $(0.5,0,l)$. Note that a linear background has been
subtracted in all scans. (d) Diagram of the $(h,0,l)$ plane to show
scan directions denoted by roman numerals.}\label{fig:fetese19:2}
\end{figure}

% dynamic susceptibility definition
The magnetic scattering cross-section is directly proportional to
the magnetic response function $S(\bm{Q},E)$ -- the Fourier
transform of the space- and time-dependent spin--spin correlation
function. According to the fluctuation--dissipation theorem,
$S(\bm{Q},E)$ is in turn related to the imaginary part of the
dynamical susceptibility \chiQw\ by \cite{shirane-book}
\begin{equation}
S(\bm{Q},E) = \frac{1}{\pi}\left[ n(E,T) + 1\right]
\chi''(\bm{Q},E).\label{eq:fluc_diss}
\end{equation}
The Bose--Einstein population factor $n(E,T)=[\exp(E/k_{\rm B} T) -
1]^{-1}$ (where $k_{\rm B}$ is the Boltzmann constant) takes into
account the increase in scattering from  bosonic excitations due to
thermal population at temperatures $T > 0$. Correction for this
factor allows the temperature dependence of \chiQw\ to be studied.

% Inelastic dispersion at (0.47,0,0.5)
Figure~\ref{fig:fetese19:2}(a) shows background corrected scans along the $(h,0,0.6)$
direction at energy transfers of 2, 4 and 6\,meV for the \feteseB\
crystal. A peak at ${\bm Q} = {\bm q}$ is present in each scan,
indicating a strongly dispersing excitation. The broadening of the
dispersion in $\bm{Q}$ may be due to unresolvable splitting of the
mode into two excitations at higher energies. The measured magnetic
response at 2\,meV parallel to $(1,0,0)$ for $l=0.5$, 0.7 and 0.9, as
shown in figure~\ref{fig:fetese19:2}(b), reveals considerable
broadening of \chiQw\ in the out-of-plane direction.  Such
broadening is characteristic of a quasi-two-dimensional system with
weak interactions along $c$. Figure~\ref{fig:fetese19:2}(c) shows
that spin fluctuations persist up to at least 150\,K, well into the
paramagnetic state. At 40\,K, i.e. close to the magnetic ordering
temperature, \chiQw\ is almost the same as at 2\,K.

\begin{figure}
\centering
\includegraphics[width=0.9\textwidth]{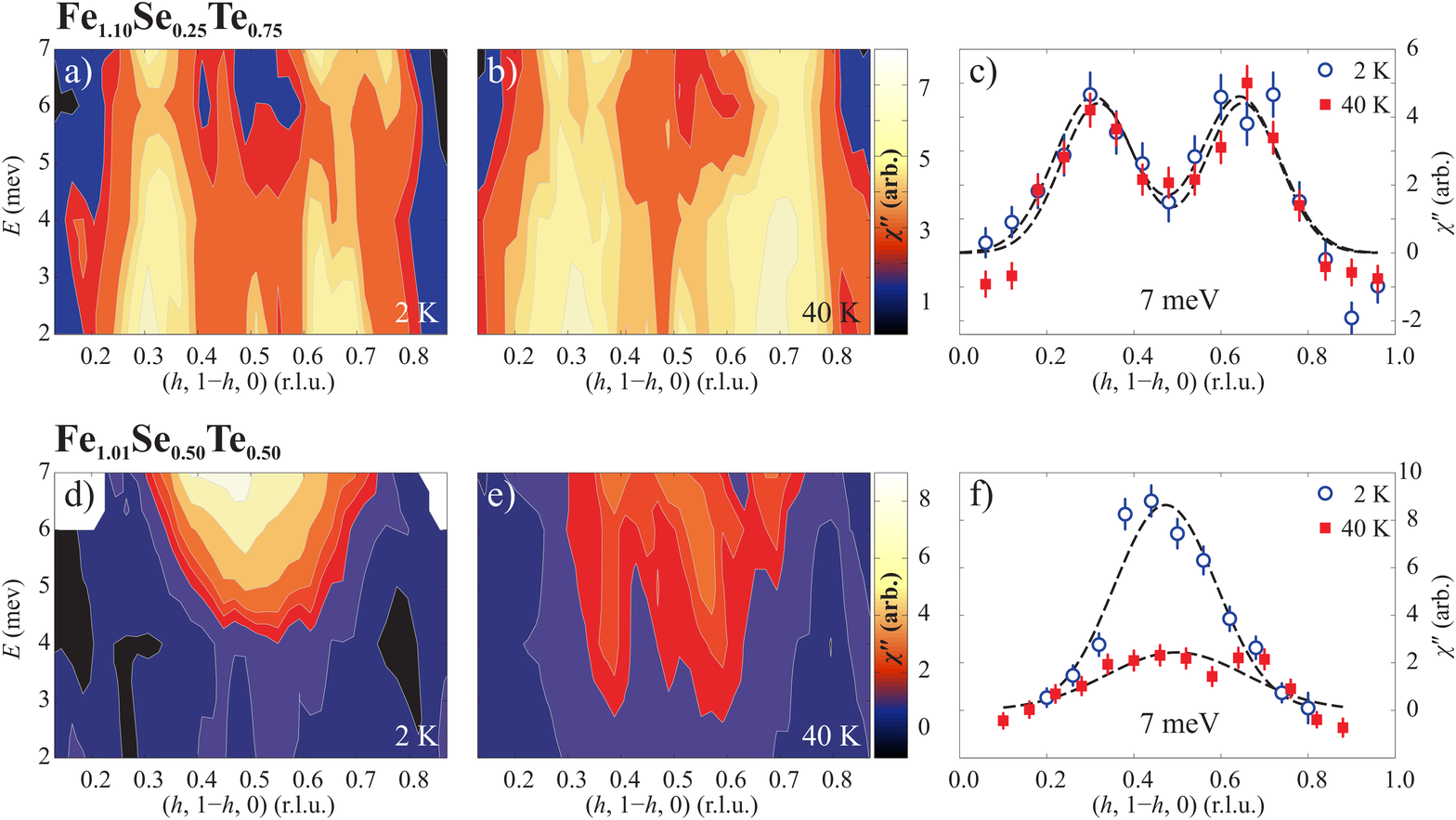}
\caption{Variation of \chiQw\ in the $(h,1-h,0)$ direction for
energies between 2\,meV and 7\,meV at temperatures of 2 and 40\,K. Data in (a)--(c) are from \feteseB\ and in (d)--(f) are from \feteseA. Constant energy cuts at 7\,meV along
$(h,1-h,0)$, measured at 2 and 40\,K for \feteseB\ and \feteseA\ are
shown in (c) and (f) respectively. A flat background has been subtracted in all scans and dashed lines through the data are fits with a Gaussian lineshape.}\label{fig:excitations}
\end{figure}

We now turn to the low energy excitation spectrum in the vicinity of
the wavevector $(1/2,1/2,0)$. Figures~\ref{fig:excitations}(a) and
(b) show maps of \chiQw\ measured along $(h,1-h,0)$ for \feteseB\ at
2 and 40\,K. The fluctuations measured at 2\,K are consistent with the magnetic excitation spectrum at higher energies reported for Fe$_{1.03}$Se$_{0.27}$Te$_{0.73}$ \cite{lumsden-nature-2010}. The excitation spectrum at 2\,K is characterised by steep
incommensurate branches arising from $(1/2\pm\epsilon
,1/2\mp\epsilon,0)$ where $\epsilon\approx 0.18$. The incommensurate excitation are still present at 40\,K. The scans shown in figure~\ref{fig:excitations}(c)
reveal that at $E = 7$\,meV, the system response is nearly the same
at 2\,K as at 40\,K. The background corrected \chiQw\ for the \feteseB\ sample does not appear to change for energies in the 2 to 7\,meV range measured at these temperatures. This is also the case for measurements along $(1/2,0,l)$ in
figure~\ref{fig:fetese19:2}(c) that show \chiQw\ data at 2\,meV to be
similar at 2 and 40\,K.

The results obtained for \feteseA\ are in stark contrast to those of
the non-superconducting \feteseB\ sample just described.
Figures~\ref{fig:excitations}(d) and \ref{fig:excitations}(e) show
maps of the magnetic spectrum as a function of wavevector along
$(h,1-h,0)$ for energies between 2\,meV and 7\,meV at 2 and 40\,K. At
2\,K we find a strong signal in \chiQw\ centred on $\bm{Q} =
(1/2,1/2,0)$ and $E \sim 7$\,meV. This feature corresponds to the
spin resonance reported previously in superconducting
FeSe$_{0.4}$Te$_{0.6}$ \cite{qui-prl-2009}, FeSe$_{0.46}$Te$_{0.54}$
\cite{argyriou-preprint-2009} and FeSe$_{0.5}$Te$_{0.5}$ \cite{mook-preprint-2009v2}. At higher energies, the excitations have
been found to disperse away from $(1/2,1/2,0)$ along $(1,-1,0)$
\cite{argyriou-preprint-2009}. However, it is the low energy
response of the system which shows the most dramatic change on
transition into the superconducting state, as may be seen in figure
\ref{fig:excitations}(f). As the sample is cooled from 40\,K to 2\,K,
the integrated dynamical susceptibility of the peak at 7\,meV increases by more than a factor of 2
and decreases in width along $(1,-1,0)$ by $\sim30$\%. Fluctuations
continue to be observed well above \Tc.

% Discussion =================================================================
\section{Discussion}

In combination with earlier measurements, the results presented here
establish that the low-energy magnetic dynamics of \fetesex\ vary
strongly with $x$. The magnetic spectra of the magnetically-ordered
compound ($x=0.25$) and the bulk superconductor ($x=0.5$) both
contain low-energy magnetic fluctuations in the vicinity of the
antiferromagnetic wavevector $(1/2,1/2,0)$. However, at $x=0.25$ the fluctuations are incommensurate with wavevector $(1/2\pm\epsilon
,1/2\mp\epsilon,0)$, $\epsilon\approx 0.18$, whereas at $x=0.5$ the
strongest magnetic signal is commensurate. Moreover, at $x=0.5$ the
magnetic spectrum has a gap of $\sim 6$\,meV and the size of the
signal just above the gap increases strongly at low temperatures.
This behaviour is consistent with the superconductivity-induced spin
resonance reported recently in bulk superconducting samples of
\fetesex\ of similar composition to ours
\cite{qui-prl-2009,iikubo-jpsj-2009,argyriou-preprint-2009,
mook-preprint-2009v2}, and also in related Fe
pnictide superconductors \cite{christianson-nature-2008,lumsden-prl-2009,chi-prl-2009,
li-prb-2009,inosov-nature-2009}.

A further difference is that the $x=0.25$ sample exhibits
short-range, static (within the \muSR\ time window) magnetic order with a characteristic wavevector
$\bm{q}=(1/2-\delta,0,\pm1/2)$, $\delta \approx 0.03$, whereas
according to our \muSR\ data there is no static magnetic order in
the bulk superconductor. The magnetic ordering wavevector $\bm q$
found at $x=0.25$ is the same as that in the parent phase
Fe$_{1+y}$Te. The slight incommensurability is thought to be caused by
the small excess of Fe accommodated in interstitial sites in the
crystal structure \cite{bao-prl-2009,fang-epl-2009,ma-prl-2009}, although it is interesting that the
incommensurability is the same to within experimental error at $y =
0.10$ (the present sample) and at $y=0.03$ (the sample studied by us
previously \cite{khasanov-prb-2009}).

Our results suggest that there are two distinct magnetic ordering
tendencies at $x=0.25$, one with wavevector $(1/2-\delta,0,\pm1/2)$
and the other with wavevector $(1/2\pm\epsilon ,1/2\mp\epsilon,0)$.
The \muSR\ data indicate that the volume fraction of magnetically
ordered phase is close to 100\%, but we cannot say whether the two
characteristic magnetic correlations coexist on an atomic scale or
whether the sample is magnetically inhomogeneous.

% Disagreement over value of TN between muons and neutrons
Finally, we comment on the fact that in the $x=0.25$ sample diffuse peaks are observed in the elastic (within energy
resolution) channel below $T \approx50$\,K by neutron
scattering but static magnetic order is only detected below $T \approx35$\,K
by \muSR. These observations can be reconciled by the difference in
the fluctuation rates observable by muons ($\sim$ GHz) and neutrons
($\sim$ THz) below which spin freezing is measured. We infer from
this that the characteristic fluctuations of the spin system lie
between $\sim$GHz and $\sim$THz for  35\,K $\lesssim T \lesssim$
50\,K. Such a gradual slowing down of the fluctuations could be a
consequence of the quasi-two-dimensional nature of the spin system,
which is also indicated by the persistence of spin correlations to
temperatures well above the ordered phase. It is also interesting that the size of the magnetically ordered domains does not significantly increase with decreasing temperature, which suggests that the short-range order is never truly static but fluctuates down to the lowest temperature investigated. This picture is consistent with the recent observation of spin-glass behaviour in Fe$_{1.1}$Se$_x$Te$_{1-x}$ for $0.05 < x < 0.55$ \cite{paulose-preprint-2009}.

% Conclusion =================================================================
\section{Conclusion}

We have observed a resonance-like peak at the
antiferromagnetic wavevector $(1/2,1/2,0)$ in the low-energy
magnetic spectrum of \feteseA, and shown that this feature is absent
from the magnetic spectrum of \feteseB\ which instead shows
incommensurate peaks flanking $(1/2,1/2,0)$. Our results reveal a
clear distinction between the magnetic excitation spectra of
\fetesex\ samples which are magnetically ordered and those which are
bulk superconductors. We conclude that the existence of a resonance
peak at the commensurate antiferromagnetic wavevector is a
characteristic of bulk superconductivity in \fetesex.

% Acknowledgments ============================================================
\ack
This work was performed at the Paul Scherrer Institut, Villigen, Switzerland. P.B. is grateful for the provision of a studentship from the U.K. Engineering and Physical Sciences Research Council.

% References =================================================================
\section*{References}
\bibliographystyle{unsrt}
\bibliography{biblio}
\end{document}